\documentclass[aps,prstab,superscriptaddress,longbibliography, twocolumn,amsmath,amssymb]{revtex4-2}	
\usepackage{graphicx}
\usepackage{dcolumn} 
\usepackage{hyperref}
\hypersetup{
    colorlinks=true,        
    linkcolor=blue,          
    citecolor=blue,        
    filecolor=magenta,      
    urlcolor=blue           
}
\newcommand{\Figref}[1]{Fig.\ref{#1}}

\newcommand{\Tabref}[1]{Table~\ref{#1}}

\newcommand{\citenamefont}{}
\newcommand{\bibnamefont}{}

\usepackage{color} 				
\usepackage[normalem]{ulem} 			

\begin{document}
\title{Acceleration and Focusing of Positron Bunch in a Dielectric Wakefield Accelerator
with Plasma in Transport Channel
}

\author{P.I. Markov}
\affiliation{NSC Kharkov Institute of Physics and Technology, 61108 Kharkov, Ukraine}
\author{R.R. Kniaziev}
\affiliation{NSC Kharkov Institute of Physics and Technology, 61108 Kharkov, Ukraine}
\author{G.V. Sotnikov}\email{sotnikov@kipt.kharkov.ua; Gennadiy.Sotnikov@gmail.com}
\affiliation{NSC Kharkov Institute of Physics and Technology, 61108 Kharkov, Ukraine}

\date{\today}

\begin{abstract}
  The paper presents the results of numerical PIC-simulation of the positron bunch focusing when accelerating in a plasma dielectric wakefield accelerator. The wakefield was excited by the drive electron bunch in a quartz dielectric tube, embedded in the cylindrical metal waveguide. The internal area of the dielectric tube has been filled with plasma having in the general case the paraxial vacuum channel. Two different models of the plasma density–radius relationship were investigated: the homogeneous model and the inhomogeneous dependence characterized the capillary discharge. Results of numerical PIC simulation have shown that it is possible a simultaneous acceleration and focusing of the test positron bunch in the wakefield. The dependence of transport and acceleration of the positron bunch with change in the size of vacuum channel, waveguide length, and the plasma density-radius model is studied.
\end{abstract}
\maketitle

\section{Introduction}
The dielectric wakefield accelerator (DWA) is considered to be a promising approach for building the TeV-energy range electron-positron colliders and compact devices in various fields of science and technology~\cite{Asman2021ARD, Shiltsev_2012, gai_power_jing_2012, Colby2010ICHEP}.

Yet, despite the possibilities of attaining high acceleration rates, demonstrated by theory~\cite{Ros1990PRD,Ng990PRD,Zha1997PRE,PhysRevSTAB.12.061302}  and experiments~\cite{Gai1988PRL,Thompson2008PRL, OShea2016, PhysRevLett.108.244801, Antipov2012APL, Shchelkunov2012PRSTAB}, one problem inherent to linear accelerators~\cite{Bal1983ICHEA-12} is not solved completely. It consists in stabilization of the transverse motion of the drive and accelerated bunches and, thus, in obtaining the accelerated particles bunches with small emittance. This DWA shortcoming can lead to beam breakup instability (BBU)~\cite{Gai1997PRE,She2008TP}. BBU instability limits the maximum accelerating gradient attainable in collinear DWA~\cite{Gai2014PRSTAB}.

For the BBU instability suppression in DWA, it has been recently proposed to use the profiled quadrupole focusing together with constant energy chirp of the drive bunch~\cite{Baturin2018PRSTAB}. This method requires very high gradients of magnetic fields; and, besides, the use of an energy chirp deteriorates the quality of the beam considering stringent requirements imposed on the beams used in colliders and free electron lasers.  The BBU instability of a bunch train can be stabilized when using resonator design of DWA~\cite{Sot2020JoI-5}. However setting up this structure in order to reach high energy of accelerated bunches is challenging.

The other way to improve the bunch transport in the DWA is to use the focusing properties of plasma medium~\cite{Ruth:1984pz,CHIADRONI201816}. We have analyzed this possibility analytically and numerically for the case of filling the bunch transport channel of DWA with plasma homogeneous in the cross section~\cite{Sot2014NIMA}. If the plasma is created as a result of a capillary discharge~\cite{Bob2001PRE,Stein2006PRSTAB}, then the transport of witness electron bunches can be even improved~\cite{Sot2017EPJ,Sot2020JoI-9}. To obtain strong focusing fields, a superdense plasma is not required~\cite{Sot2020JoI-9}. Experimental studies of the transport of electron bunches in plasma-filled wake dielectric accelerators (PDWA) have already started~\cite{Berezina2016UFZh,BIAGIONI2018247}, and the first experimental confirmation of the focusing of electron bunches in the 2.8~GHz DWA has been obtained~\cite{Berezina2016UFZh,Oni2016PAST-PP}. It should be noted that the focusing of accelerated witness bunch electrons is observed for certain conditions on the plasma density~\cite{Sot2014NIMA,Sot2020JoI-9}.

Until recently, all studies of the bunch transport in the vacuum DWA have been related to the electron bunches. In principle, the transport of positron bunches should not differ qualitatively from the transport of electron ones. However, using of plasma in the transport channel can make it difficult to transport the positron bunches. The positron witness bunch transport in PDWA may face the same problems that exist in beam-driven plasma wakefield accelerator studies (see~\cite{LebedevPRAB2017,Diederichs_PRAB2019,Diederichs_PRAB2020} and references there). Although the first analytical studies have shown the possibility of focusing the accelerated positron bunches in PDWA~\cite{Kniaziev2013PAST-PE,Sot2014NIMA}, a full numerical simulation, that takes into account the self-consistent dynamics of the drive electron and accelerated positron bunches with due regard to the group velocity effects~\cite{Bal2001JETP}, has started but recently. In the paper~\cite{Mar2021PAST-PE} we have investigated the dependence of energy and diameter of the accelerated positron bunch versus the diameter of a paraxial vacuum channel at various plasma density-radius relation values. At these simulations the delay time of witness positron bunch had remained invariable with change in the vacuum channel diameter. Our studies have demonstrated that despite the fact that simultaneous acceleration of the positron bunch by the dielectric wakefield component of the electron bunch and its focusing by the plasma wave does take place, the positron bunch energy characteristics were not satisfactory because of a wide energy spread. Further studies at other delay times have resulted in significant improvement for both the energy characteristics and the focusing of the positron bunch~\cite{sotnikov:ipac2021-mopab145}. However, keeping the positron-bunch delay time constant relative to the time of drive electron bunch injection with change of the vacuum channel diameter precluded reaching the highest energy of the accelerated positrons.

Preliminary results of the delay time optimization of the accelerated positrons have been carried out in~\cite{MarkovArXiv2021_v1} for various values of the paraxial vacuum channel diameters with a homogeneous distribution of the plasma density in it. These results show the significant energy gain increase of the accelerated positron bunch without deteriorating its focusing in comparison with the non-optimized case~\cite{sotnikov:ipac2021-mopab145}.

In present paper the optimization on the delay time for the acceleration and transport of the positron bunch by the drive electron bunch wakefield was continued. In addition to the case presented in~\cite{MarkovArXiv2021_v1}, we considered the case of an inhomogeneous distribution of the plasma density in a transport channel with a transverse profile, which is realized in a capillary discharge~\cite{Bob2001PRE}.

\section{The problem definition}
In our researches the dielectric tube of internal radius $a$  and outer radius  $b$ , iwhich was inserted into a cylindrical metal waveguide was used. The internal region of the dielectric tube between radii $r_{p1}$  and $a$  was filled with plasma. Thus, within the system there was the paraxial cylindrical vacuum column of radius  $r_{p1}$.

\begin{figure}
  \centering
  \includegraphics[width=0.5\textwidth]{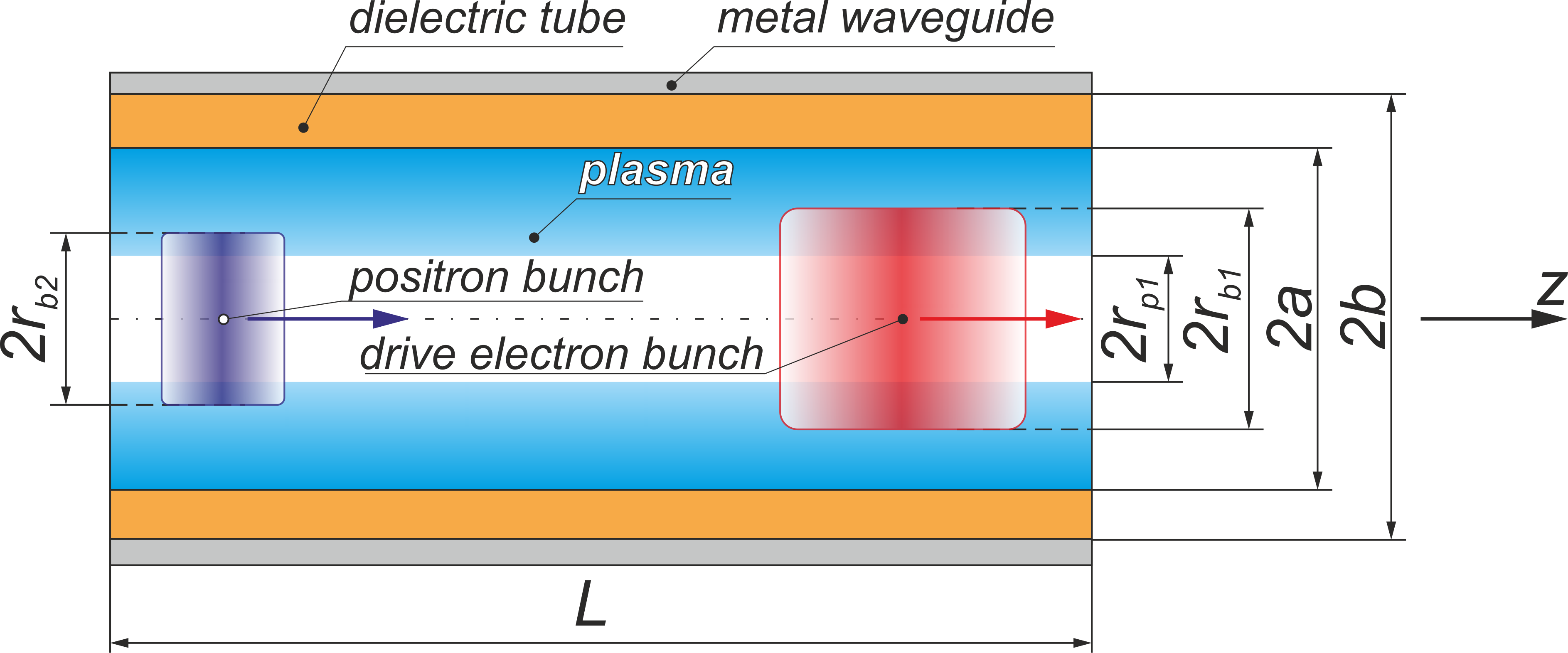}
  \caption{Schematic view of longitudinal section of the plasma dielectric wakefield accelerator of positrons.}\label{Fig:01}
\end{figure}

The cylindrically shaped drive electron bunch of radius  $r_{b1}$  passed through the slowing-down structure along its axis and excited a wakefield. In a given delay time $t_{del}$ following the drive bunch the positron bunch with the absolute value of charge lower by a factor of 60 than that of the drive bunch, was injected in the system along its axis.  The radius of positron bunch was  $r_{b2}$. We call the plasma-filled structure with the drive electron bunch and the witness positron bunch as the plasma dielectric wakefield accelerator of positrons (PDWAP).  Its longitudinal section is presented schematically in~\Figref{Fig:01}, where the pink cylinder shows the drive electron bunch, the violet cylinder shows the positron bunch. The blue and orange colors represent the plasma-filled region and the dielectric tube, respectively.

The parameters, used in calculations are specified in~\Tabref{tab:01}.

\begin{ruledtabular}
\begin{table}
\centering
\caption{The parameters of the waveguide, drive and witness (test) bunches, used in calculations of PDWAP.}\label{tab:01}
\begin{tabular}{|l|r|}
Inner dielectric-tube radius $a$             & $0.5\,mm$          \\
Outer dielectric-tube radius $b$             & $0.6\,mm$          \\
Inner plasma-cylinder radius $r_{p1}$        & $0\div 0.5\,mm$    \\
Waveguide length, $L$                        & 8\,mm              \\
Dielectric permittivity $\varepsilon$        & $3.75 (quarz)$     \\
Bunch energy $E_0$                           & $5\,GeV$           \\
Drive electron bunch charge                  & $-3\,nC$           \\
Witness positron bunch charge                & $0.05\,nC$         \\
Longitudinal rms deviation of drive bunch          &              \\
charge $2\sigma_1$(Gauss charge distribution) & $0.1\,mm$         \\
Longitudinal rms deviation of positron bunch          &           \\
charge $2\sigma_2$(Gauss charge distribution) & $0.05\,mm$        \\
Total drive bunch length                     &                    \\
used in PIC simulation                       & $0.2\,mm$          \\
Total positron bunch length                  &                    \\
used in PIC simulation                       & $0.1\,mm$          \\
Drive bunch diameter $2r_{b1}$               & $0.9\,mm$          \\
Positron bunch diameter $2r_{b2}$            & $0.7\,mm$          \\
Paraxial plasma density $r_{p1} = 0$         & $2\cdot 10^{14}\,cm^{-3}$ \\
\end{tabular}
\end{table}
\end{ruledtabular}

\begin{figure}
  \centering
  \includegraphics[width=0.5\textwidth]{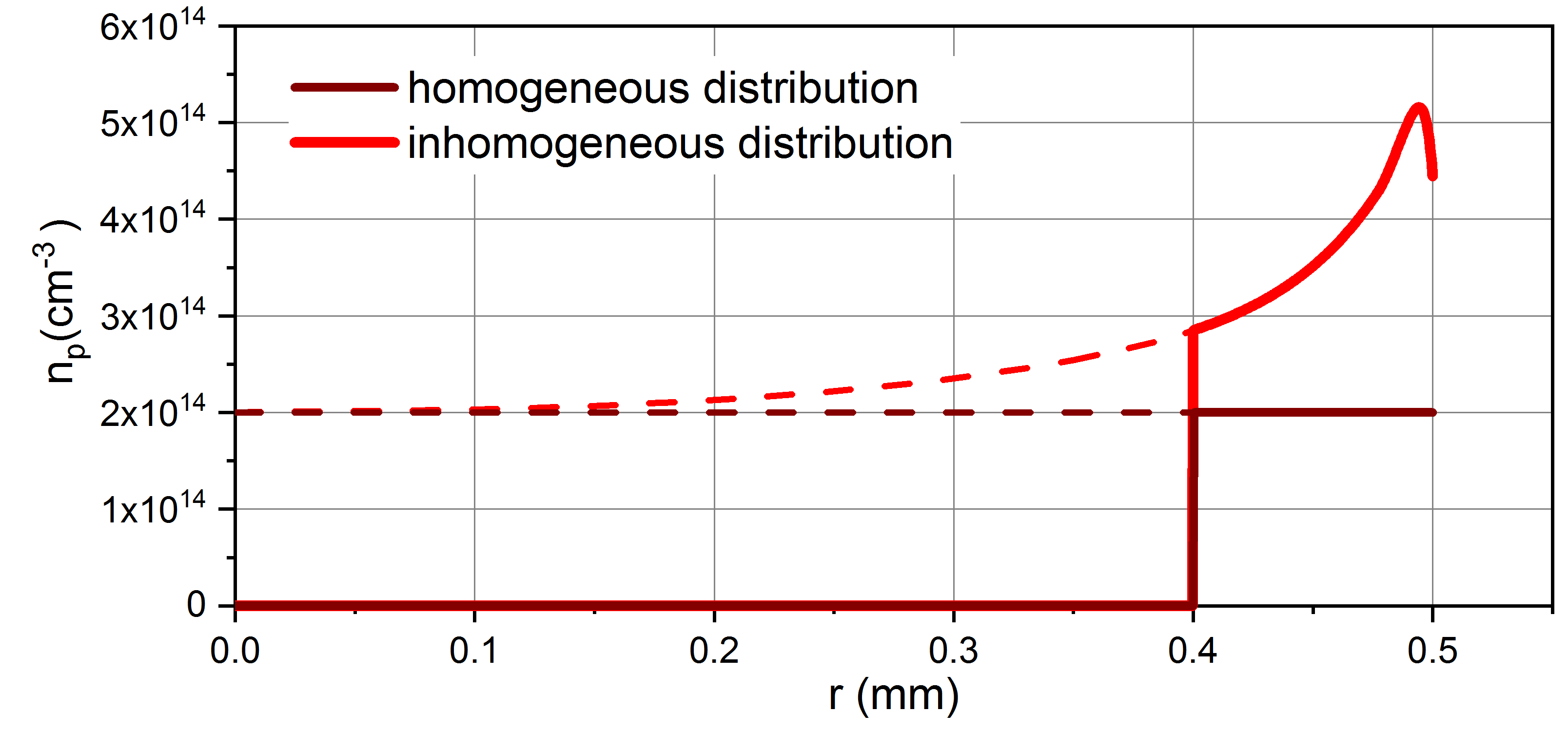}
  \caption{Models of plasma density-radius relationship $n_p(r)$  for two cases: the plasma cylinder fully filling the interior of the dielectric tube (dotted line) and the plasma cylinder of internal radius  $r_{p1}=0.4\,mm$ (solid line). Red line corresponds to the inhomogeneous distribution of plasma density, brown line is for the homogeneous one.}\label{Fig:02}
\end{figure}

When simulating the capillary discharge two different models of the plasma density-radius relationship $n_p(r)$ were investigated: 1) the homogeneous model and 2) the inhomogeneous dependence, characterized the capillary discharge~\cite{Bob2001PRE}. At the vacuum-plasma boundary at  $r=r_{p1}$ the stepwise behavior of $n_p(r)$  as a functions of radius $r$  was assumed, (see~\Figref{Fig:02}).

\section{Technique of optimum delay time finding}

\begin{figure*}[!tbh]
  \centering
  \includegraphics[width=0.75\textwidth]{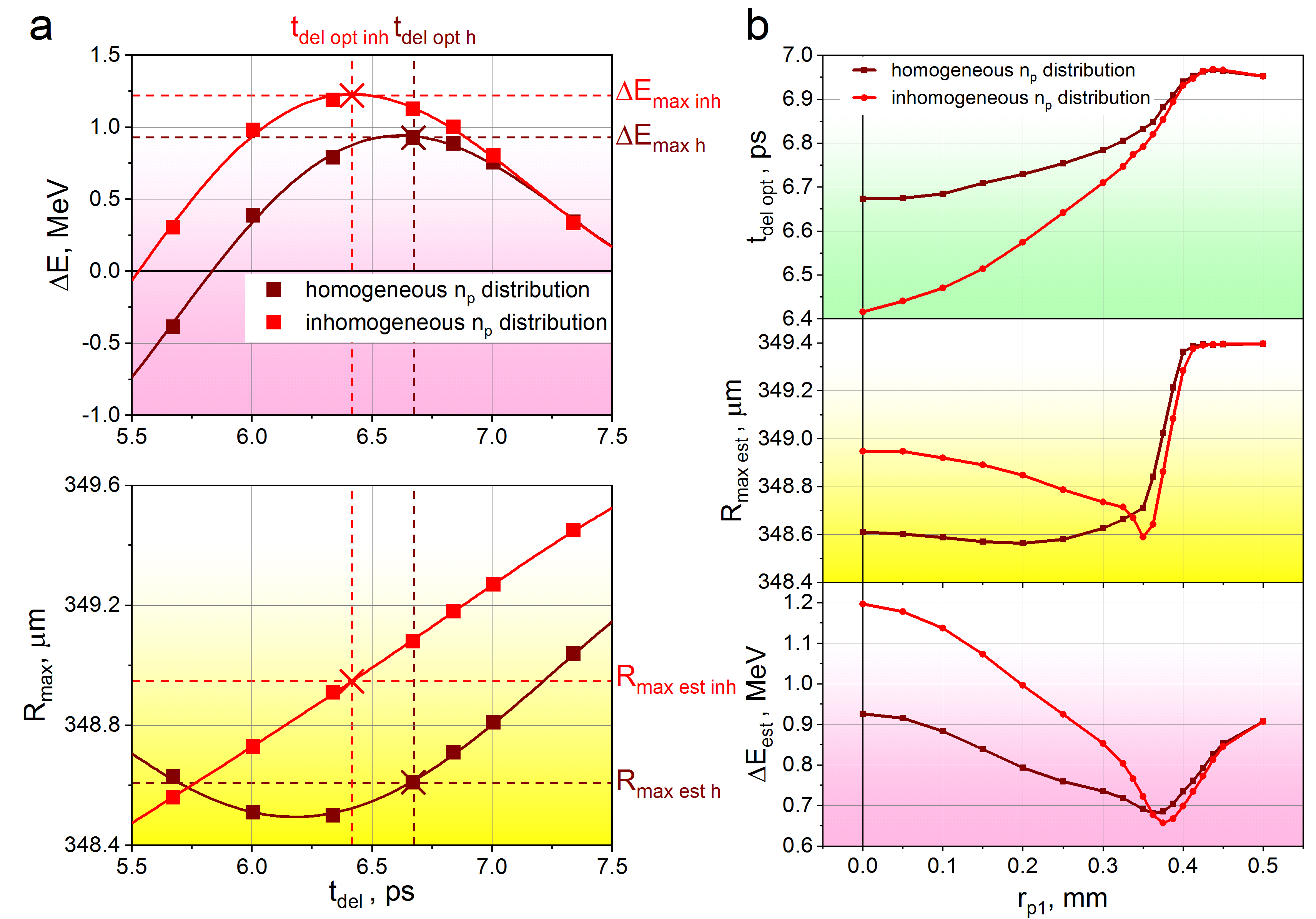}\hfill
  \caption{a) The test bunch energy gain dependence and the expected radius on delay time; the best values $t_{del,opt}$ of test bunch and the radius corresponding to this time $R_{max\;est}$  are shown by crisscross; $r_{p1}=0$. Polynomial dependences are shown by solid lines. b) Dependence of optimum delay time values $t_{del,opt}(r_{p1})$ (at the top), the expected behavior of the test bunch radii $R_{max\;est}(r_{p1})$ (in the middle) and the estimated energy gain in the positron bunch  $\Delta E_{est}(r_{p1})$ (at the bottom).}\label{Fig:03}
\end{figure*}
As was stated above, for obtaining the greatest possible average energy of the accelerated positron bunch it is necessary to calculate the optimum delay time value of the positron bunch injection relative to the drive electron bunch  $t_{del}$  with due regard for the vacuum column radius  $r_{p1}$ value. Finding of optimum $t_{del}$  value was carried out as follows. For the specified vacuum channel radius $r_{p1}$  the wakefield excitation by the drive bunch and the dynamic motion of the accelerated positron bunch in the field were simulated at several close  $t_{del}$  values. Each time with that, the average energy gain $\Delta E$ of the positron bunch was defined. Based on the obtained  $\Delta E(t_{del})$   values the polynomial passing through the points  $(t_{del,i},\Delta E_i)$  was calculated and its maximum  $\Delta E_{max}$  was defined. The abscissa of maximum point was just the optimum $t_{del,opt}$  value.

\Figref{Fig:03}a a exemplifies finding $t_{del,opt}$ and the expected test bunch radius $R_{max\;est}$   that corresponds to $t_{del,opt}$    found for the case of the dielectric tube internal region fully filled with plasma $r_{p1}=0$. The calculations for an array of radii in the range $0\le r_{p1}\le a$  have resulted in obtaining the optimum delay time values $t_{del,opt}$, the expected behavior of test bunch radii $R_{max\;est}$   and the estimated gain in the positron bunch energy  $\Delta E_{est}$ in the form of $r_{p1}$  functions shown in~\Figref{Fig:03}b, where the brown color shows the values relating to the homogeneous plasma density in the dielectric tube channel model, red --- to the $n_p(r)$  at the capillary discharge.

\section{Results of 2.5-dimensional PIC-simulation}

In numerical simulation by means of our own 2.5D PIC code~\cite{Sot2014NIMA}  we studied the wakefield topography and also, the dynamics of electron and positron bunches when moving in the drift chamber.  For the paraxial vacuum column effect analysis, we have investigated several variants with the different initial inner plasma cylinder radii $r_{p1}$  varying in the  $0\le r_{p1} \le a$ range.

\begin{figure*}
  \centering
  \includegraphics[width=\textwidth]{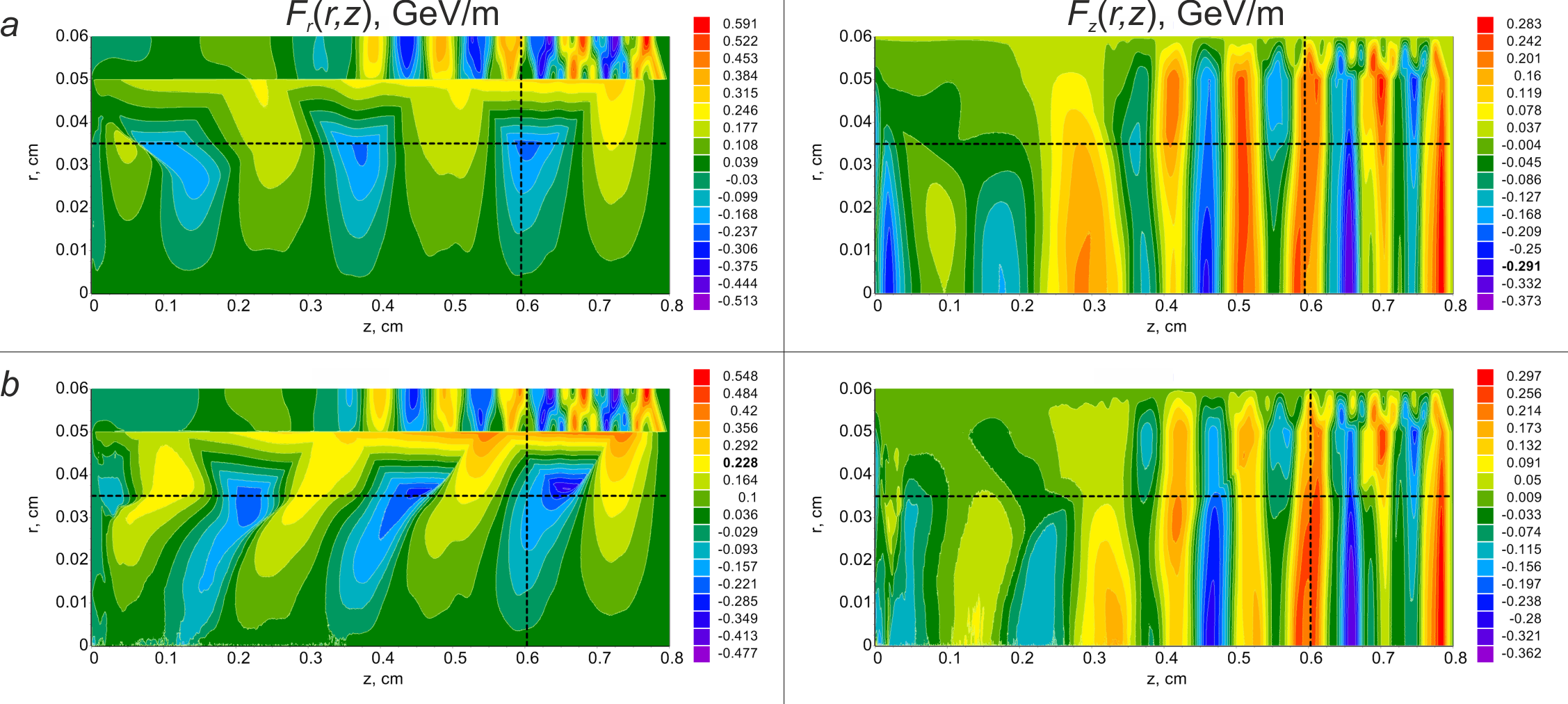}
  \caption{Color maps and level lines for transverse $F_r(r,z)$  (at the left) and longitudinal $F_z(r,z)$ (at the right) components of Lorentz force, acting on test positron for time $t=26.69\,ps$ at different plasma density-radius relationships: a) the homogeneous model, b) the dependence realized in capillary discharge for $r_{p1}=0$  case.}\label{Fig:04}
\end{figure*}

\Figref{Fig:04} shows comparative pictures of the Lorentz force components acting on the test positron in the PDWAP for $t=26.69\,ps$ at different plasma density-radius relationships a) the homogeneous,  b) the inhomogeneous dependence realized in capillary discharge for the case $r_{p1}=0$ (that is, for continuous filling of the drift channel with plasma). The vertical dashed line shows the test bunch position.  The horizontal dotted line corresponds to $r=r_{b2}$. In the place where the test bunch is positioned, the $F_r$ value is negative (shown in various shades of cold colors), but $F_z$ is positive (various shades of warm colors), that should lead to an acceleration of the bunch positrons with their simultaneous focusing.

\begin{figure*}
  \centering
  \includegraphics[width=\textwidth]{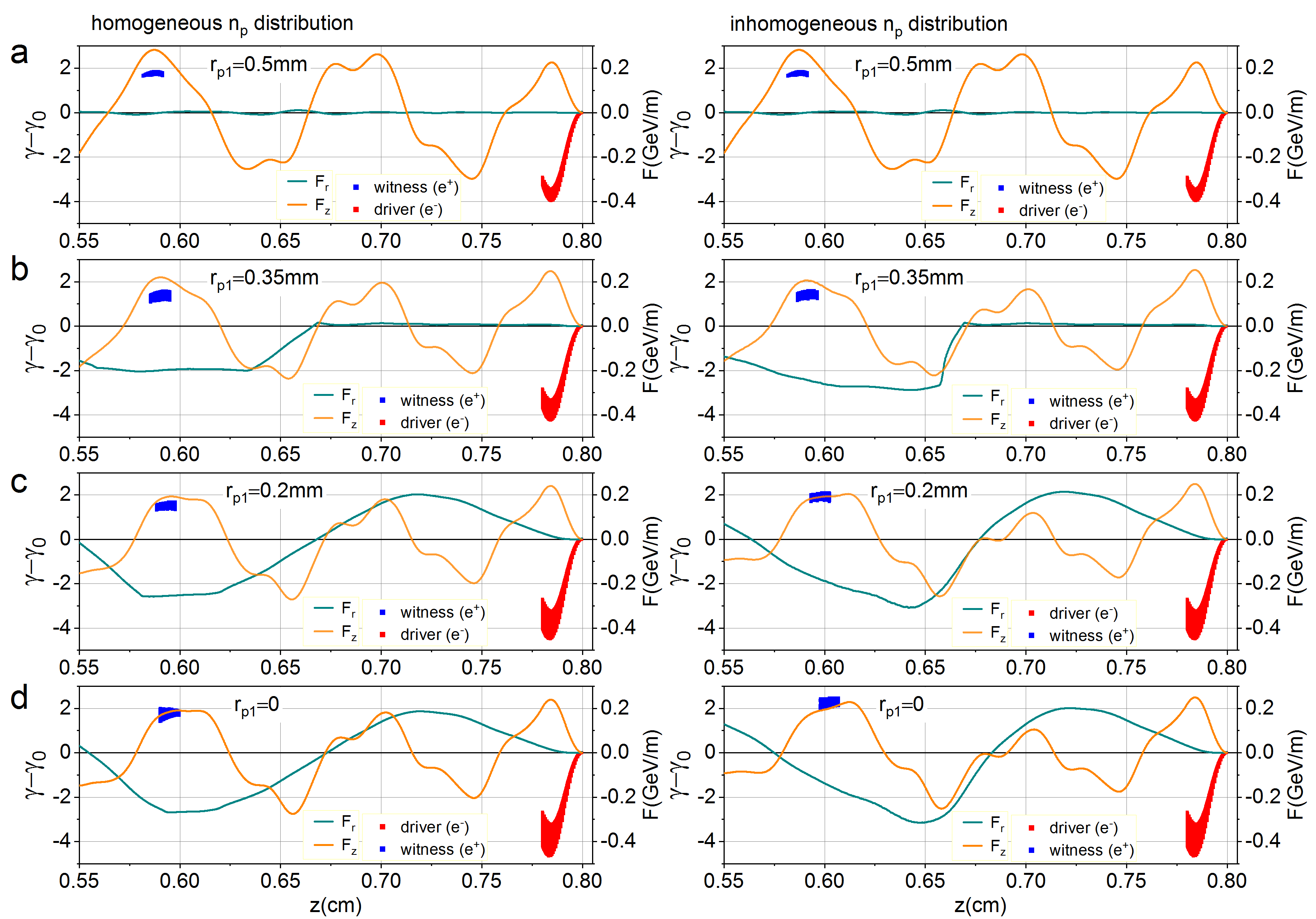}
  \caption{Phase space energy vs longitudinal coordinate, combined with longitudinal  $F_z(z)$  and transverse  $F_r(z)$  force functions at
$r = {r_{b2}}$  for time  $t=26.69\,ps$   at different $r_{p1}$   values: a) $r_{p1}=0.5\,mm$, b) $r_{p1}=0.35\,mm$, c) $r_{p1}=0.2\,mm$   and d) $r_{p1}=0$. Blue color cloud shows positrons energy of the accelerated bunch, red — electrons energy of the drive bunch.}\label{Fig:05}
\end{figure*}

For more obvious illustration of axial vacuum tube radius $r_{p1}$  effect on test bunch focusing and acceleration, \Figref{Fig:05} shows the phase space in combination with longitudinal  $F_z(z)$  and transverse  $F_r(z)$  force functions at  $r=r_{b2}$ (that is the test bunch radius) for the same time as that in~\Figref{Fig:04}, at different $r_{p1}$   values: a) $r_{p1}=0.5\,mm$, b) $r_{p1}=0.35\,mm$, c) $r_{p1}=0.2\,mm$   and d) $r_{p1}=0$. Note that the a) case in~\Figref{Fig:05} corresponds to lack of plasma in the drift channel,  and the d) case --- to its full filling with plasma.  The red color near output end of the phase space shows the electron energy of the drive bunch. The blue color shows the positron energy of accelerated bunch. The left column represents the homogeneous plasma distribution, and the right column --- the inhomogeneous plasma distribution.
As is evident from the plot in~\Figref{Fig:05}a, in the absence of plasma in the drift channel, the transverse force $F_r$  is negligible. As a result, the test bunch focusing is also practically absent.  As the radius  $r=r_{p1}$  is reduced, the thickness of the plasma cylinder interacting with the test bunch increases, and increasing part of the positron bunch penetrates the plasma region. \Figref{Fig:05}b, c and d illustrate the cases $r_{p1}=0.5\,mm$, $r_{p1}=0.35\,mm$, $r_{p1}=0.2\,mm$  and $r_{p1}=0.0$,  respectively. It can be seen that therewith the transverse force $F_r(z)$   arises too. Its minimum (i.e. maximum of focusing) is located practically in the same place, where the test positron bunch is positioned.  The $F_r$  value influencing the test bunch ranges within $-(0.202\div 0.196) GeV/m$, $-(0.257\div 0.253) GeV/m$  and $-(0.251\div 0.268) GeV/m$, respectively for the homogeneous plasma case; and within $ - \left( {0.233 \div 0.254} \right)\,{{{\rm{GeV}}} \mathord{\left/  {\vphantom {{{\rm{GeV}}} {\rm{m}}}} \right.
 \kern-\nulldelimiterspace} {\rm{m}}}$,
$ - \left( {0.164 \div 0.200} \right)\,{{{\rm{GeV}}} \mathord{\left/
 {\vphantom {{{\rm{GeV}}} {\rm{m}}}} \right.
 \kern-\nulldelimiterspace} {\rm{m}}}$ and
 $ - \left( {0.127 \div 0.173} \right)\,{{{\rm{GeV}}} \mathord{\left/
 {\vphantom {{{\rm{GeV}}} {\rm{m}}}} \right.
 \kern-\nulldelimiterspace} {\rm{m}}}$ for the case of the inhomogeneous plasma of capillary discharge. This results in focusing of the test bunch. The best positron bunch focusing is observed at ${r_{p1}} = 0.2\;{\rm{mm}}$ for homogeneous $n_p(r)$ distribution and at ${r_{p1}} = 0.35\;{\rm{mm}}$ for the $n_p(r)$ realized at capillary discharge.
\begin{figure*}
  \centering
  \includegraphics[width=0.75\textwidth]{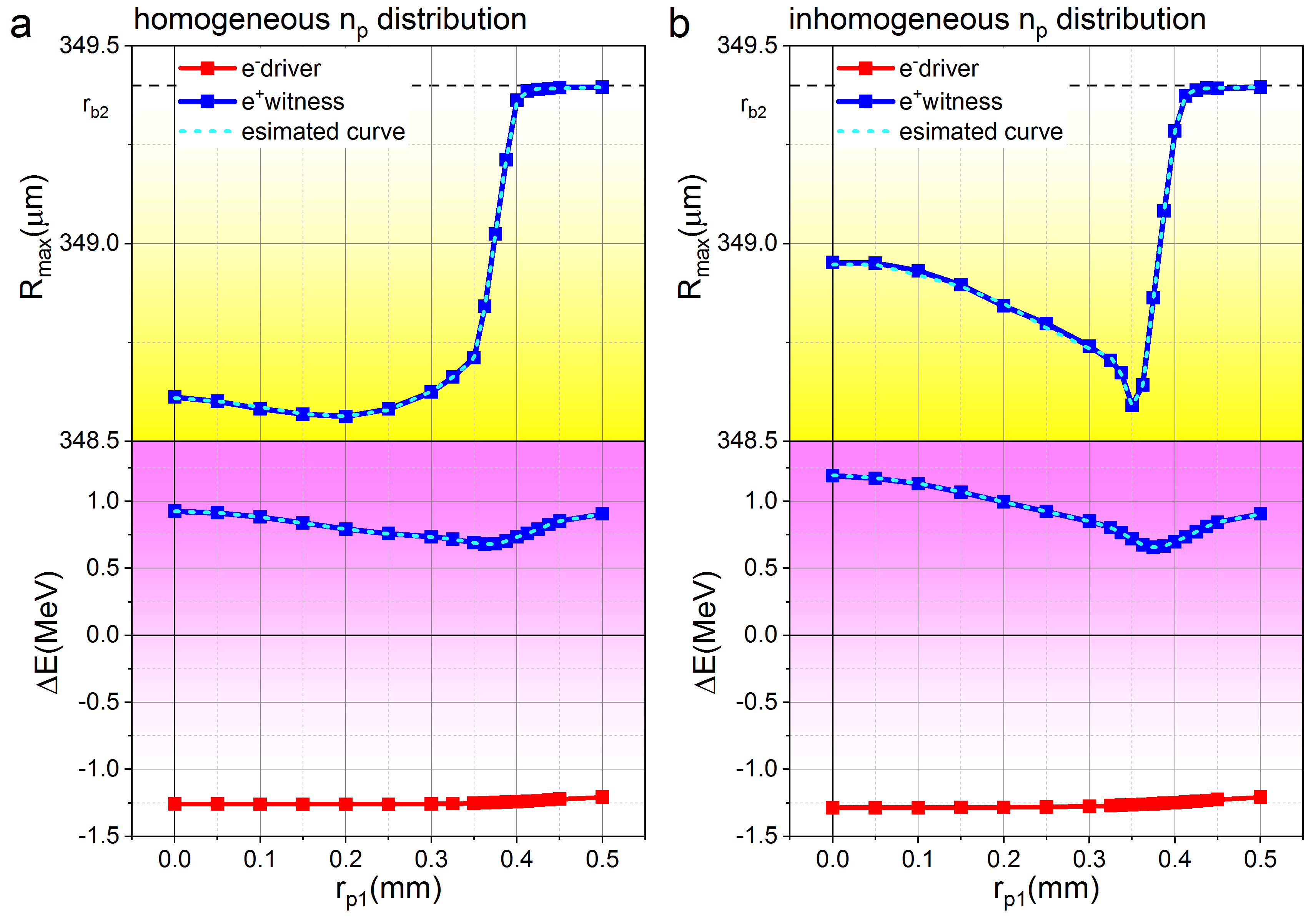}
  \caption{The test bunch radius $R_{max}$  (at the top) and energy gain  $\Delta E$  of the test bunch and the slowed-down drive bunch (at the bottom) with change in the inner plasma tube radius  $r_{p1}$  for time  $t=26.69\,ps$; a) the homogeneous plasma distribution, b) the plasma distribution dependence realized in capillary discharge.}\label{Fig:06}
\end{figure*}

Dependence of test bunch radius at change of inner plasma tube radius from $0$ to $0.5\,mm$ for time $t=26.69\,ps$ is shown in~\Figref{Fig:06} (at the top). As appears from the curves shown in~\Figref{Fig:06}a for the homogeneous plasma distribution at  $r_{p1}$ increase from $0$ to $0.2\,mm$  the gradual increase in the test positron bunch focusing is observed. On further $r_{p1}$ increase till $0.35\,mm$  (a positron bunch radius), the focusing goes down a little. Further increase in $r_{p1}$ leads to a drastic degradation in the test bunch focusing. If $r_{p1}\ge 0.425\,mm$ the test bunch focusing is practically absent. In case of inhomogeneous transverse profile of plasma density a gradual increase in the test positron bunch focusing is observed with an $r_{p1}$ increase increase from $0$ to $0.35\,mm$ (see~\Figref{Fig:06}b).

 \Figref{Fig:06}(from below) shows the energy gain  $\Delta E$  of the test bunch (blue curve) and the slowed-down drive bunch (red curve) versus the inner plasma-tube radius $r_{p1}$  change.  With $r_{p1}$  increase from $0$ to $0.375\,mm$, a small energy gain  $\Delta E$   reduction of the test bunch is observed. The further  $r_{p1}$  increase from $0.375\,mm$ to $0.5\,mm$ leads to $\Delta E$  of the test bunch increasing. When using a homogeneous plasma, an interesting result should be noted: the increase in the energy of a positron bunch in the vacuum case practically coincides with the increase in energy in the case of the complete filling of the transport channel with plasma.

The blue dotted line in \Figref{Fig:06} has shows the expected  $R_{max\;est}$ (corresponding to the found $t_{del,opt}$  values) and $\Delta E_{est}$  values obtained when deriving the optimum delay time  $t_{del,opt}$  by the above-mentioned polynomial interpolation method.  Almost complete agreement between the expected curves and the ones calculated by direct 2.5D-simulation gives evidence of the correctness of the chosen optimization method.

\begin{figure*}
  \centering
  \includegraphics[width=\textwidth]{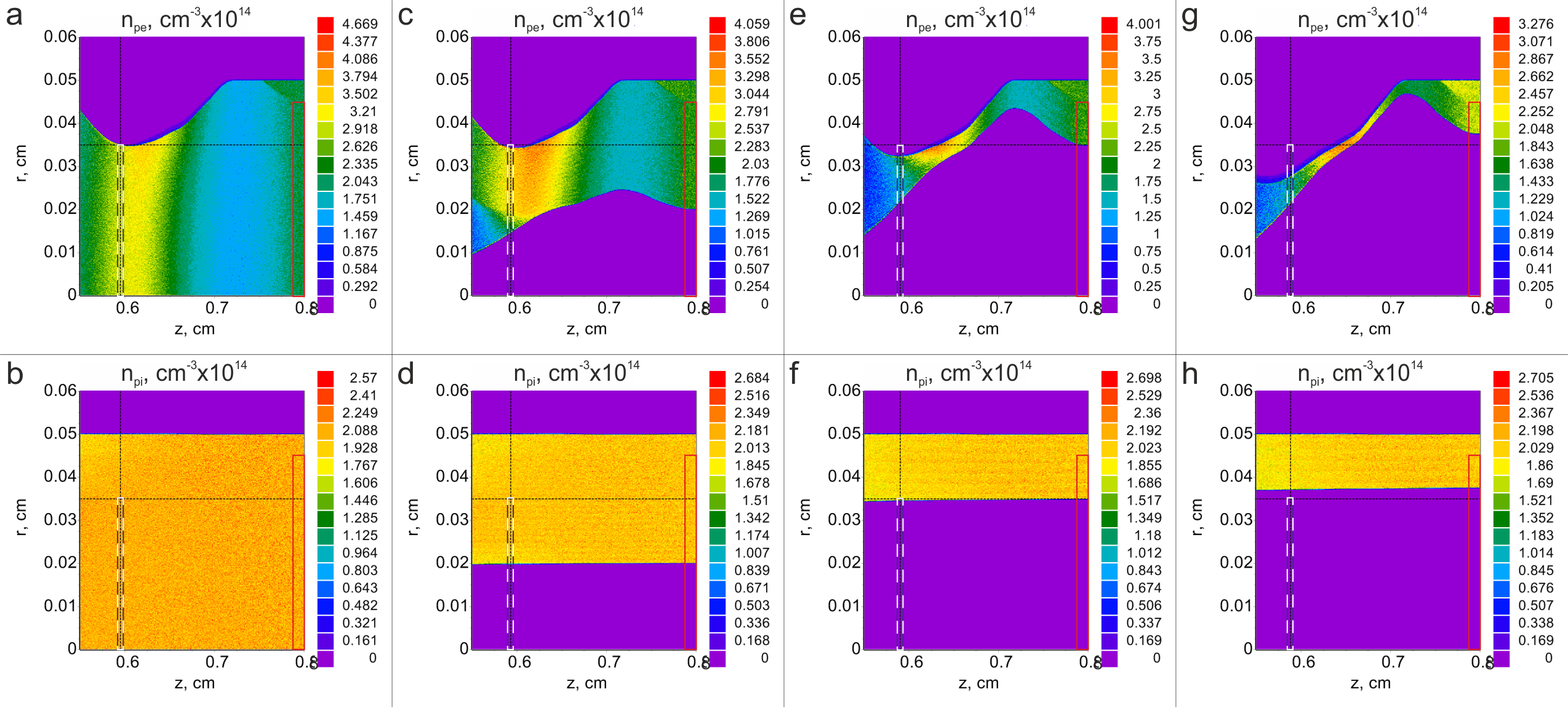}
  \caption{The plasma electron density (in the top figures) and plasma ion density (in the bottom figures) for   $r_{p1}=0$(a, b),   $r_{p1}=0.2\,mm$(c, d),   $r_{p1}=0.35\,mm$ (e, f) and   $ r_{p1}=0.375\,mm$ (g, h). Red and blue-white rectangles show the position of drive and test bunches. Plasma is homogeneous in transverse section.}\label{Fig:07}
\end{figure*}

For explanation of the test bunch transverse size behavior shown in~\Figref{Fig:06}, we consider the behavior of plasma electrons in the drift channel. Figures~\ref{Fig:07}~a, c, e and g show plasma electron density  $n_{pe}(r,z)$  for the time  $t=26.69\,ps$   for $r_{p1}=0$ , $r_{p1}=0.2\,mm$, $r_{p1}=0.35\,mm$ and $r_{p1}=0.375\,mm$, respectively. Figures~\ref{Fig:07}~b, d, f and h depict the corresponding plasma ion densities $n_{pi}(r,z)$ .

As is evident from the graphs in~\Figref{Fig:07}, the drive bunch electrons push out the plasma electrons to the periphery of the drift channel. This results in the excess of plasma ions behind the drive bunch. These ions attract the plasma electrons which were pushed out by the driver and the last ones start moving to the waveguide axis. Here it should be noted that the plasma ion density remains almost invariable during the delay time ${t_{del}}$  of the test bunch. The mentioned processes lead to formation of region having the excessive plasma electron density below the test positron bunch surface. The excess of electrons pulls positrons towards the axis. Additionally the excessive plasma ions located above the positron bunch push positrons in the same direction.

As can be seen in~\Figref{Fig:07}a, in case of full plasma filling of the drift channel, the plasma electrons appear directly inside the test bunch. In the vicinity of test bunch the plasma electrons density has a maximum. These electrons together with plasma ions (see \Figref{Fig:07}b) are favorable for causing all the test bunch positrons be affected by the force that would drive them to the waveguide axis. The presence of ion background above the test bunch surface and the excess of plasma electrons over plasma ions directly above the test bunch surface cause the test bunch focusing.

If there is a vacuum channel of radius  $r_{p1}=0.2\,mm$, all plasma electrons are also inside the test bunch in that place where the test bunch is positioned, (see \Figref{Fig:07}c). Over test bunch there are only plasma ions. Under the test bunch surface the plasma electrons density is partially compensated by the ion density. This case differs from the previous one in that not all positrons move to the axis but only those that are in the region $0.15\;{\rm{mm}} \le r \le {r_{b2}}$. However, this is quite sufficient for the test bunch focusing.t

Increasing the radius of the vacuum cylinder to  $r_{p1}=0.35\,mm$   causes an increase in the gap between the test bunch surface and the level to which the plasma electrons go down under the bunch (about $0.025\,mm$) (see \Figref{Fig:07}e). The plasma ions are over the test bunch(see \Figref{Fig:07}f). Thus, the ions push positrons to the axis and focus the test bunch. However, because the plasma electrons are far from the bunch surface and cannot affect the positrons located at the periphery, the focusing turns out to be weaker than with the smaller $r_{p1}$  values, described above.

\Figref{Fig:07}g and \Figref{Fig:07}h illustrate the case, where the plasma ions are above the test bunch surface and therefore their effect on the change in the bunch radius is small. The plasma electrons are deeply in the test bunch. Their effect on the change in radial motion of positrons located at the bunch periphery is weak. This results in poor focusing of the test bunch at $r_{p1}=0.375\,mm$.

\begin{figure*}
  \centering
  \includegraphics[width=\textwidth]{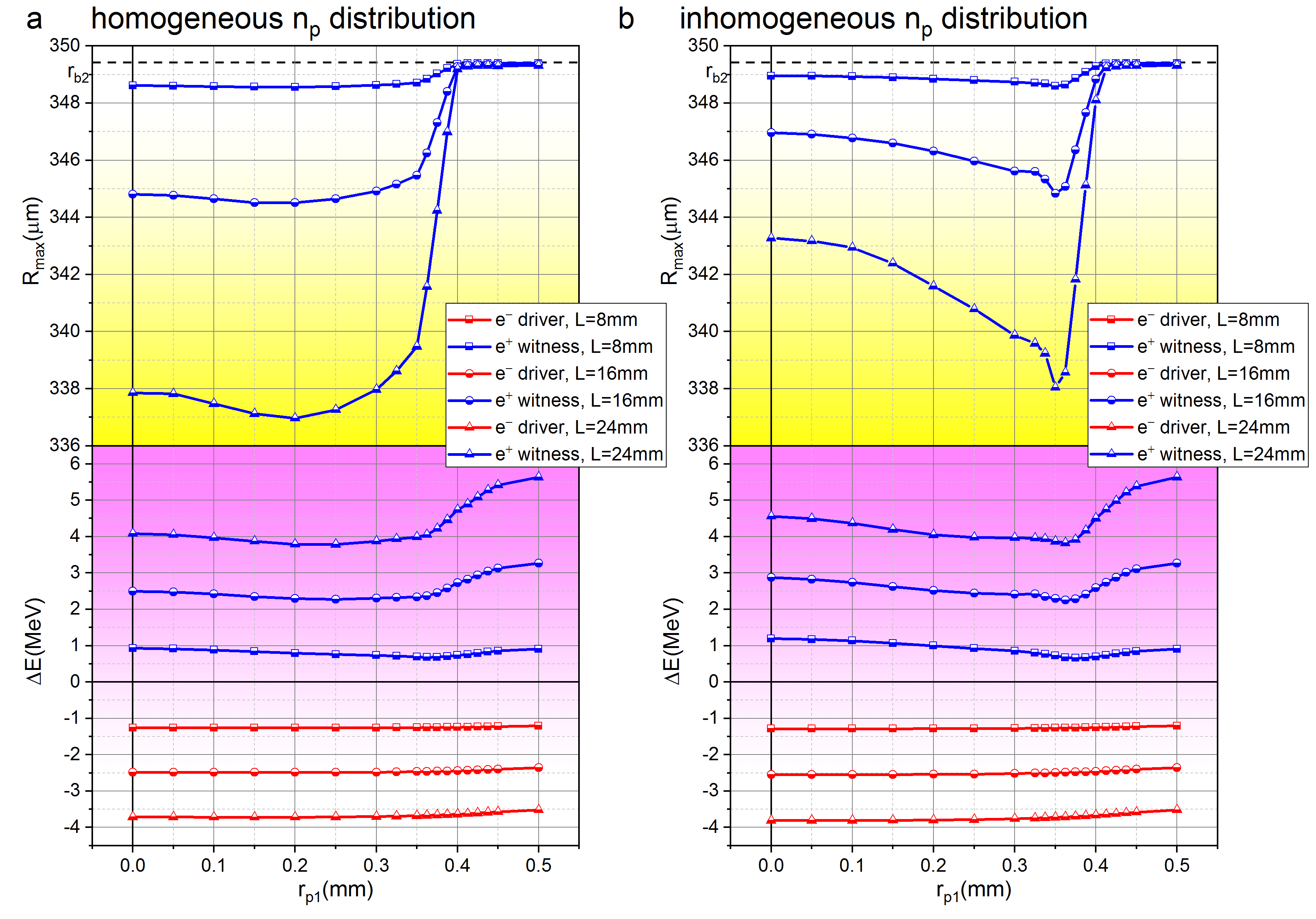}
  \caption{Test bunch radius $R_{max}$  (at the top) and energy gain  $\Delta E$  of the accelerated test bunch positrons and the slowed-down drive bunch electrons (below) with variation in the smaller plasma-tube radius $r_{p1}$  for the waveguide lengths  $L=8\,mm$,  $L=16\,mm$ and $L=24\,mm$ at $t=26.69\,ps$,  $t=53.38\,ps$ and $t=80.07\,ps$, respectively;  a) the homogeneous plasma density distribution, b) the inhomogeneous transverse plasma density profile created at capillary discharge.}\label{Fig:08}
\end{figure*}

Our study has demonstrated the feasibility of simultaneous acceleration and focusing of the positron bunch using the wakefield excited by the drive electron bunch. However, the acceleration and focusing values are very small, being no more than 0.019\% (energy gain) and 0.24\% (reduction of positron bunch diameter), respectively, for the parameters given in~\Tabref{tab:01}. The energy gain and improvement in focusing of the test positron bunch can be attained by increasing the length of the structure with keeping the other parameters unchanged. In~\Figref{Fig:08} shows the influence of waveguide length: ($8\,mm$, $16\,mm$ and $24\,mm$) on the behavior of both the test bunch radius ${R_{{\rm{max}}}}$  (at the top) and the energy gain $\Delta E$  of the accelerated positron bunch and the slowed-down electron drive bunch (bottom) with the change in the vacuum channel radius $r_{p1}$  for the time $t$  values $26.69\,ps$, $53.38\,ps$ and $80.07\,ps$ (each corresponding to the time for the drive bunch to reach the structure end). Figures~\ref{Fig:08}~a and b show, respectively, the homogeneous plasma density distribution and the plasma density distribution at capillary discharge. Based on the curves shown in~\Figref{Fig:08} it can be concluded that the qualitative behavior of the change in the energy of the driver electron and the test positron bunches and the focusing of the test bunch with the lengthening of the accelerating structure is preserved. It allows extrapolating the received results on real, longer accelerating structures.

\begin{figure*}
  \centering
  \includegraphics[width=\textwidth]{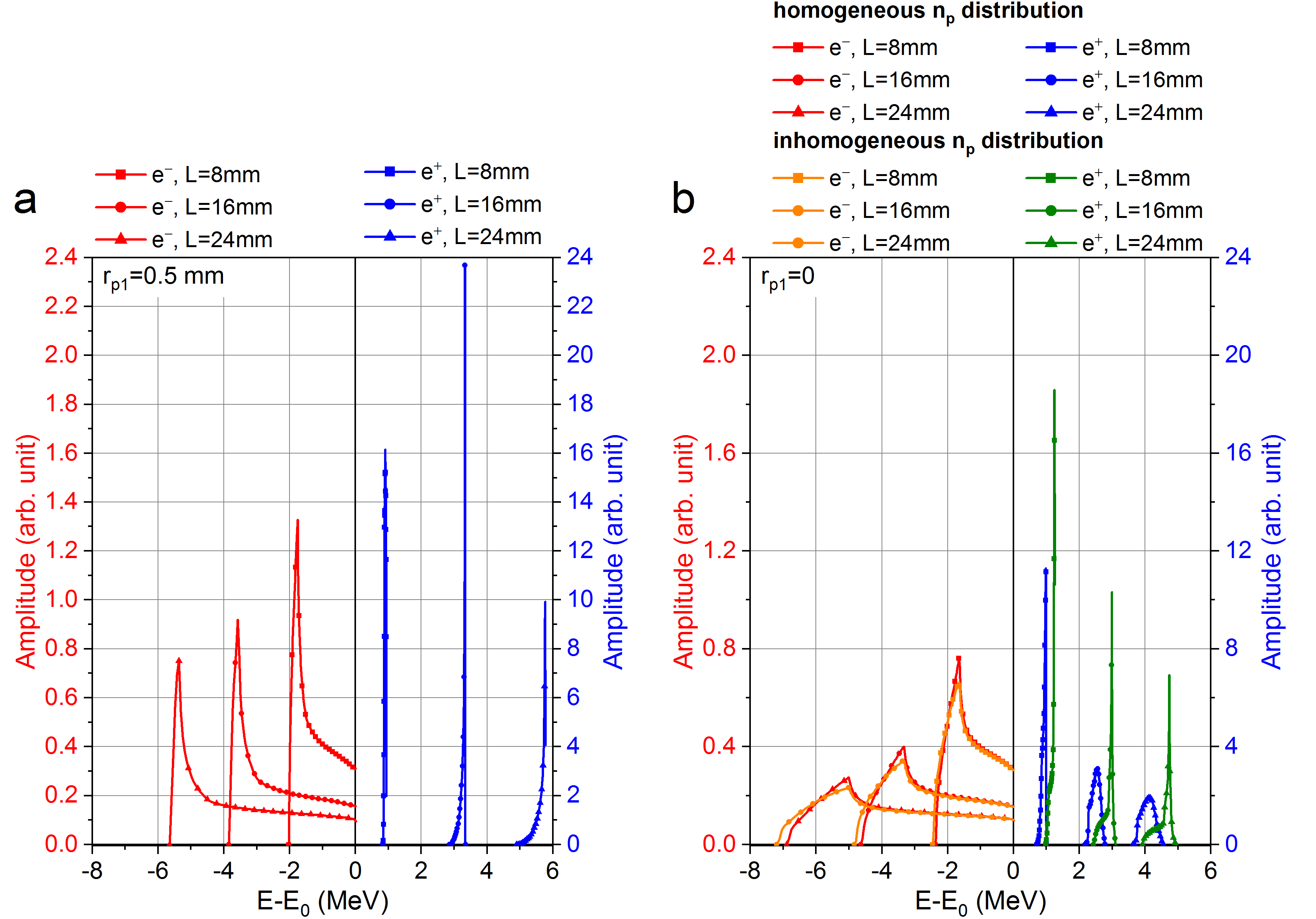}
  \caption{Energy distribution of drive electrons and witness positron bunches for times $t=26.69\,ps$, $t=53.38\,ps$ and $t=80.07\,ps$ for waveguide length  $L=8\,mm$, $L=16\,mm$ and $L=24\,mm$, respectively. The $r_{p1}$ values are a -- $0.5\,mm$, b -- $0$.}\label{Fig:09}
\end{figure*}

More detailed information on the drive-to-test bunch energy transfer is given in~\Figref{Fig:09}, which shows the energy distributions of the driver electrons and test bunch positrons for the times of $t=26.69\,ps$, $t=53.38\,ps$ and $t=80.07\,ps$ for the waveguides length  $L=8\,mm$, $L=16\,mm$ and $L=24,mm$, respectively, at different  $r_{p1}$ values. \Figref{Fig:09}a corresponds to the plasma absence in the drift channel, and \Figref{Fig:09}d --- to complete filling of the latter with plasma. It can be seen that although the injected bunches were monoenergetic, the resulting distribution has expanded. At the same time, the distribution of drive bunches is close to half-normal and test bunches --- to Gaussian-like. The lowest energy spread of the bunches is observed in the absence of plasma (\Figref{Fig:09}a). In the presence of plasma in the drift channel, the plasma inhomogeneity leads to energy increase of the test bunch, as is evident from the comparison of the blue and green curves in~\Figref{Fig:09}b.

\section{Conclusions}

Here we have presented the results of numerical PIC-simulation of wake-field excitation and self-consistent dynamics of charged particles in the plasma-dielectric cylindrical slowing-down THz-band structure in terms of the homogeneous plasma model, and the model of inhomogeneous plasma produced by capillary discharge in a waveguide.

The performed numerical simulation has confirmed the predictions of the analytical theory, demonstrated the acceleration of test positron bunch along with its simultaneous focusing.

It has also been shown that the vacuum channel in the plasma column improves the accelerated positron bunch focusing, yet reduces the bunch acceleration. The vacuum channel has the optimum radius value that provides the best focusing. In homogeneous plasma column, as the vacuum channel radius exceeds its optimum value, the focusing goes down at first smoothly, but as soon as the plasma tube surrounds the test positron bunch, fast focusing degradation occurs. In the case of inhomogeneous plasma filling, there is no region showing a smooth decrease in focusing, and the best test bunch focusing results when the inner plasma tube radius is equal to the test bunch radius.

The presence of inhomogeneous plasma in the drift channel is preferable to the homogeneous case, since it provides higher energy gain of the accelerated test bunch.

The best acceleration takes place if there is no plasma in the drift channel, however then there will be no test positron bunch focusing.

\begin{acknowledgments}
The study is supported by the National Research Foundation of Ukraine under the program “Leading and Young Scientists Research Support” (project \# 2020.02/0299).
\end{acknowledgments}

\bibliography{Bibliography}	
\end{document}